\documentclass[12pt]{article}
\usepackage{amsfonts}
\usepackage{amssymb}
\usepackage{graphicx}
\usepackage[english]{babel}
\renewcommand{\theequation}{\arabic{section}.\arabic{equation}}

\def\be{\begin{equation}}
\def\ee{\end{equation}}
\def\bea{\begin{eqnarray}}
\def\eea{\end{eqnarray}}

\def\g{\gamma}

\begin{document}

\input epsf

\begin{flushright}  hep-th/0409067
\end{flushright}
\vspace{20mm}
\begin{center} {\LARGE Geometry of D1-D5-P bound states}
\\
\vspace{20mm} {\bf  Stefano Giusto and Samir D. Mathur}\\
\vspace{4mm}

Department of Physics,\\ The Ohio State
University,\\ Columbus, OH 43210, USA\\
\vspace{4mm}
\end{center}
\vspace{10mm}
\begin{abstract}

Supersymmetric solutions of 6-d supergravity (with two translation 
symmetries) can be written as a hyperkahler base times a 2-D fiber. The 
subset of these solutions which correspond to true bound states of 
D1-D5-P charges give microstates of the 3-charge extremal black hole. 
To understand the characteristics shared by the bound states we 
decompose known bound state geometries into base-fiber form. The axial 
symmetry of the solutions make the base Gibbons-Hawking. We find the 
base to be  actually `pseudo-hyperkahler': The signature changes from 
(4,0) to (0,4) across a hypersurface. 2-charge D1-D5 geometries are 
characterized by a `central curve' $S^1$; the analogue for 3-charge 
appears to be a hypersurface that for our metrics is an orbifold of 
$S^1\times S^3$.

\end{abstract}
\thispagestyle{empty}
\newpage
\setcounter{page}{1}
\section{Introduction}\setcounter{equation}{0}

The black hole information puzzle \cite{hawking} suggests that we lack
a key factor in
our understanding of large dense collections of matter.
Some computations in string theory suggest that the traditional picture
of a
black hole as `empty space with a central singularity' might be
incorrect, and the degrees of freedom accounting for the black hole
entropy are distributed throughout the hole. The simplest object with
entropy is the 2-charge extremal system, which can be realized as a
bound state of D1 and D5 branes.
In this case it was found that the microstates are not pointlike but
have a certain typical `size'. Different microstates have different
geometries; none has a horizon or singularity but if we draw a surface
bounding the typical state then the area $A$ of this surface satisfies
${A\over 4G}\sim S$ where $S=2\sqrt{2}\pi \sqrt{n_1n_5}$ is the entropy
of the system \cite{lm4,lm5}.

We can wonder if a similar `swelling up' would happen for the 3-charge
extremal D1-D5-P states. The microscopic entropy
of the system is $S_{micro}=2\pi\sqrt{n_1n_5n_p}$. This time the naive
geometry is a black hole whose horizon area satisfies $S_{Bek}\equiv
{A\over 4G}=S_{micro}$ \cite{stromvafa}. If the microstates `swelled
up' to fill this horizon then there would be no information puzzle --
the black hole microstates would just be large `fuzzballs' that
radiated like any other ball of matter.
No individual state would have a horizon; the horizon would only be an
effective construct arising upon coarse graining over microstates.

In \cite{gms1,gms2,lunin} the geometries for some 3-charge microstates
were constructed, and it was found that
these geometries were regular, with no horizon. The `throat' of the
3-charge solution, instead of ending in a horizon,
ended in a `cap', just like the 2-charge geometries. But these
geometries were a small subset of all the states of the 3-charge
system, and were also not very generic states; in particular they
carried a significant amount of rotation.

The generic state is not expected to be well-described by a classical
geometry. But in the 2-charge case the classical geometries helped
deduce the size of the generic microstate, and we expect to get
significant insight
from classical geometries in the 3-charge case as well.  The extremal
solutions we seek are BPS and thus the geometries preserve
supersymmetry. The  D1-D5-P system is obtained by compactifying IIB
string theory to $M_{4,1}\times S^1\times T^4$. Dimensionally reducing
on $T^4$, we get supersymmetric solutions in 6-D. In the classical
limit of large charges we expect translation invariance along the time
directions $t$ and the $S^1$ direction $y$ \cite{lm4}.

In \cite{gmr} the general class of supersymmetric 6-D supergravity
geometries (with these translation symmetries) was described. All such
solutions
can be written as a 4-di\-men\-sion\-al hyperkahler base with a 2-dimensional
$(t,y)$ fiber over this base. In an interesting set of recent papers,
such formulations have been used to construct large families of
3-charge BPS solutions \cite{bw,emparan3,gg, bk2}.

But generic solutions constructed this way include regular ones as well
as ones with pathologies  (horizons, singularities, closed timelike
curves). They include true bound states of D1-D5-P charges  as well as
superpositions of such bound states. Unlike the situation with D1-D5 ,
we do not have as yet a way to isolate the solutions that describe the
true bound states
of D1-D5-P, which are the ones that describe the microstates of the
3-charge black hole.

To get some insight into the characteristics of  bound states, in
this paper we return to the geometries constructed in
\cite{gms1,gms2}. These are known to describe true bound states, since
they were constructed by starting with 2-charge
D1-D5
bound states and doing spectral flow to add the P charge; a further
class was obtained by applying S,T dualities to these geometries so
that we again get  bound states.  We cast these solutions in the form
of \cite{gmr}, identifying the base and the fiber. From the result we
can
immediately make some observations. A hyperkahler metric is usually
Euclidean, with signature
$(4,0)$. We find that the base for our geometries is actually
`pseudo-hyperkahler': The signature on the outer region is $(4,0)$ but
inside a certain boundary it changes to $(0,4)$.\footnote{The full 6-D 
metric however retains signature
$(5,1)$ everywhere.}

The place where the signature changes is given by the intersection of a
surface $f=0$ in 6-D with the 4-D base. In the 2-charge D1-D5 case the
$f=0$ condition defined the `central curve' of the KK-monopole tube
which characterized the geometry: Different shapes of this curve gave
different bound states. This motivates us to study the $f=0$ surface in
the 3-charge case as well. We find that this surface (at $t=const.$)
is an orbifold of $S^3\times S^1$ by an
orbifold group $Z_k$; the group acts without fixed points so the $f=0$
surface is smooth. When the P charge vanishes the $S^3$ collapses to
zero and we get the $S^1$ of the 2-charge D1-D5 system.

The geometries of \cite{gms1, gms2} are regular
everywhere except for possible  ALE type singularities arising from
fixed points of an orbifold action.  In the above base-fiber split the
metric of the base turns out to have certain orbifold singularities,
and we investigate how the fiber at these locations
behaves so as to yield the singularities of the full 6-D metric. We end
with some
conjectures on the role of the $f=0$ surface: Since different shapes of
the $f=0$ curve described all different bound states of the 2-charge
D1-D5 system, it might be that different shapes of the
$f=0$ surface in the 3-charge case characterize all different D1-D5-P
bound states.

\section{Writing the metrics as base $\times$ fiber}

Consider Type IIB string theory on $M_{5,1}\times S^1\times T^4$. Let
the length of $S^1$ be $2\pi R$ and the volume of $T^4$ be $V$. Wrap
$n_5$ D5 branes on $S^1\times T^4$ and $n_1$ D1 branes on $S^1$. In
addition let there be $n_p$ units of momentum along $S^1$. We are
interested in the geometries created by the BPS bound states of these
charges.

Dimensionally reducing on $T^4$ we get supersymmetric solutions in 6-D.
In our solutions we can choose moduli such that
the solution can be represented as a solution of minimal supergravity
in 6-D (in particular the dilaton becomes  a constant). Further, we
expect that in the classical limit which we consider the solutions will
be translationally invariant in the time direction $t$ and the $S^1$
direction $y$ \cite{lm4}.

\subsection{General form of the 6-D metrics}

It was shown in \cite{gmr} that the most general supersymmetric
solution to minimal supergravity in 6 dimensions
with translation symmetry along
\be
u=t+y\,,\quad v = t-y
\ee
can be written as
\be
ds^2= -H^{-1}(dv + \sqrt{2} \beta)\Bigl(du + \sqrt{2} \omega + {F\over
2}(dv + \sqrt{2} \beta)\Bigr) + H h_{mn} dx^m dx^n
\label{general6d}
\ee
Here $x^m$ ($m=1,\ldots,4$) are coordinates in the noncompact spatial
directions.  $H$ and $F$ are functions and $\beta$ and $\omega$ are
1-forms on this 4-dimensional space.   $h_{mn} dx^m dx^n$ gives a
hyperkahler metric on this 4-dimensional space; we call this the `base
metric'.

To write the field equations satisfied by these variables, define the
self-dual 2-form on the base
\be
\mathcal{G}^+= H^{-1}\,\Bigl({d\omega + \star d\omega\over 2}+{F\over
2}\,d\beta\Bigr)
\label{g+}
\ee
where the $\star$ operation is defined with respect to the hyperkahler
base
metric.
Then the equations of motion are equivalent to the following non-linear
system of equations for
$H$, $F$, $\beta$ and $\omega$
\bea
d\beta&=&\star d\beta\label{beta}\\
d\mathcal{G}^+&=&0\label{g}\\
d\star d\,H+d\beta\,\mathcal{G}^+&=&0\label{h}\\
d\star d\,F +\Bigl(\mathcal{G}^+\Bigr)^2 &=&0\label{f}
\eea

\subsection{Writing 3-charge solutions in the general form}

In \cite{gms1,gms2} certain solutions of IIB supergravity were
constructed carrying 3 charges, $Q_1$, $Q_5$, $Q_p$, and 2 angular
momenta
(parametrized by $\gamma_1,\gamma_2$). When the $Q_1$ and $Q_5$
charges are set equal ($Q_1=Q_5=Q$), and the moduli at infinity chosen
appropriately,
the dilaton vanishes and the 3-charge solution reduces to a solution of
minimal supergravity. The resulting 6-D metric is
\bea
\label{3chargemetric}
ds^2 & = & -\frac{1}{h} (dt^2-dy^2) + \frac{Q_{p}}{h
f}\left(dt-dy\right)^{2}+ h f \left( \frac{dr^2}{r^2 +
(\g_1+\g_2)^2\eta} + d\theta^2
\right)\nonumber \\
          &+& h \Bigl( r^2 + \g_1\,(\g_1+\g_2)\,\eta -
\frac{Q^2\,(\g_1^2-\g_2^2)\,\eta\,\cos^2\theta}{h^2 f^2}
\Bigr)
\cos^2\theta d\psi^2  \nonumber \\
&+& h\Bigl( r^2 + \g_2\,(\g_1+\g_2)\,\eta +
\frac{Q^2\,(\g_1^2-\g_2^2)\,\eta\,\sin^2\theta}{h^{2} f^{2}
}
\Bigr) \sin^2\theta d\phi^2  \nonumber \\
&+& \frac{Q_p\,(\g_1+\g_2)^2\,\eta^2}{h f}
\left( \cos^2\theta d\psi + \sin^2\theta d\phi \right)^{2} \nonumber\\
&-& \frac{2 Q }{hf}
\left(\g_1 \cos^2\theta d\psi + \g_2 \sin^2\theta d\phi\right)
(dt-dy)
\nonumber \\
&-& \frac{2 Q\,(\g_1+\g_2)\,\eta}{h f}
\left( \cos^2\theta d\psi + \sin^2\theta d\phi \right) dy
\eea
with
\bea
&&\!\!\!\!\!\!\!\!\!\!\!\!Q_p=-\g_1\g_2\,,\quad \eta = {Q\over Q + 2
Q_p}\nonumber\\
&&\!\!\!\!\!\!\!\!\!\!\!\!f = r^2+ (\g_1+\g_2)\,\eta\,
\bigl(\g_1\, \sin^2\theta + \g_2\,\cos^2\theta\bigr) \nonumber\\
&&\!\!\!\!\!\!\!\!\!\!\!\!h =1+\frac{Q}{f}
\label{deffh}
\eea
These geometries are dual to specific microstates of the D1-D5-P
system. Thus the angular momenta $\gamma_1, \gamma_2$ take specific
values; for these values  the geometry (\ref{3chargemetric}) has
no horizon, no closed time-like curves and the only
singularities are orbifold singularities (which can
be understood as degenerations of smooth geometries). The values of
$\gamma_1, \gamma_2$  fall into two discrete series.
Let
\be
a={Q\over R}
\ee
($R$ is the radius of the $y$ circle).  The first series is
\be
\g_1 = - a\,n\,,\quad \g_2=a\,\Bigl(n+{1\over k}\Bigr), \quad
n, k\in\mathbb{Z}
\label{sf}
\ee
The second series corresponds to geometries obtained from the first by
    S,T dualities which interchange the D1 and P charges. These 
geometries
have
\be
\g_1=-a \,{k\over n_5 (k n+1)}\,\quad \g_2=a\,{1\over n_5 n},\quad
n, k \in\mathbb{Z},~~~n_5\in\mathbb{N}
\label{st}
\ee

The metric (\ref{3chargemetric}) can be written in the form
(\ref{general6d}), with the following values for
$H$, $F$, $\beta$, $\omega$ and $h_{mn}$
\bea
H&=&h\nonumber\\
{F\over 2}&=& -{Q_p\over f}\nonumber\\
\sqrt{2}\beta&=& {Q\over f}\,(\g_1 + \g_2)\,\eta\,(\cos^2\theta\,d\psi
+ \sin^2\theta\,d\phi) \nonumber\\
\sqrt{2}\omega&=& {Q\over f}\,\Bigl[\Bigl(2\g_1 -
(\g_1 + \g_2)\,\eta\,\Bigl(1-2 {Q_p\over
f}\Bigr)\Bigr)\,\cos^2\theta\,d\psi \nonumber\\
&&\qquad\qquad\qquad + \Bigl(2\g_2 - (\g_1 + \g_2)\,\eta\,\Bigl(1-2
{Q_p\over f}\Bigr)\Bigr)\,\sin^2\theta\,d\phi\Bigr]
\label{3chargegmr}
\eea
\bea
&&\!\!\!\!\!\!\!\!\!\!\!\!h_{mn} dx^m dx^n= f\Bigl({dr^2\over r^2+(\g_1
+ \g_2)^2\, \eta} + d\theta^2\Bigr)\nonumber\\
&&\!\!\!\!\!\!\!\!\!\!\!\!\qquad\quad + {1\over f}\Bigl[[r^4 +
r^2\,(\g_1+\g_2)\,\eta\,(2\g_1 - (\g_1-\g_2) \cos^2\theta) +
(\g_1+\g_2)^2\,\g_1^2\,\eta^2\,\sin^2\theta]\cos^2\theta\,d\psi^2
\nonumber\\
&&\!\!\!\!\!\!\!\!\!\!\!\!\qquad\quad\quad\,\,+[r^4 +
r^2\,(\g_1+\g_2)\,\eta\,(2\g_2 + (\g_1-\g_2) \sin^2\theta) +
(\g_1+\g_2)^2 \,\g_2^2 \,\eta^2 \,\cos^2\theta]\sin^2\theta
\,d\phi^2\nonumber\\
&&\!\!\!\!\!\!\!\!\!\!\!\!\qquad\quad\quad\,\,-2\g_1 \g_2
\,(\g_1+\g_2)^2\, \eta^2\,\sin^2\theta \cos^2\theta \,d\psi d\phi\Bigr]
\label{3chargebase}
\eea

\subsection{The base metric in Gibbons-Hawking form}

Since the geometry (\ref{3chargemetric}) satisfies the equations of
motion, the base metric
     $h_{mn} dx^m dx^n$  (eq. \ref{3chargebase}) should turn out to be
hyperkahler.
      In fact this base metric is more special -- it has two commuting
isometries
corresponding to translations along  $\phi$ and $\psi$.
Then, according to a theorem of \cite{gibbons}, at least one linear
combination of the two isometries is tri-holomorphic (i.e. commutes
with the 3 complex structures of the hyperkahler space) and $h_{mn}
dx^m dx^n$ should  be writable in Gibbons-Hawking form
\be
h_{mn} dx^m dx^n=H_2^{-1}(d\tau + \chi \,d{\tilde \phi})^2 + H_2\,ds_3^2
\label{gh}
\ee
Here $\tau$ and ${\tilde \phi}$ are linear combinations of $\phi$ and
$\psi$,
$ds_3^2$ is the flat metric on $\mathbb{R}^3$. $H_2$ is a function on
$\mathbb{R}^3$ harmonic in the metric
$ds_3^2$ and $\chi\,d{\tilde \phi}$ is a 1-form on $\mathbb{R}^3$ that
satisfies
\be
dH_2 = \star_3 \,d(\chi\, d{\tilde\phi})
\ee
($\star_3$ is the Hodge dual with respect to the metric $ds_3^2$).

We can cast  (\ref{3chargebase}) in the form (\ref{gh}) if we choose
\be
\tau=\psi-\phi\,,\quad \tilde\phi=\psi+\phi
\ee
Then one finds, starting from (\ref{3chargebase}),
\be
H_2={4\,f\over (r^2 + (\g_1+\g_2)^2\,\eta\,\cos^2\theta)\,(r^2 +
(\g_1+\g_2)^2\,\eta\,\sin^2\theta)}
\ee
\be
\chi={\g_1\over
\g_1+\g_2}\,{r^2\,\cos\,2\theta+(\g_1+\g_2)^2\,\eta\,\cos^2\theta\over
r^2 + (\g_1+\g_2)^2\,\eta\,\cos^2\theta}+{\g_2\over
\g_1+\g_2}\,{r^2\,\cos\,2\theta-(\g_1+\g_2)^2\,\eta\,\sin^2\theta\over
r^2 + (\g_1+\g_2)^2\,\eta\,\sin^2\theta}
\ee
and
\bea
ds_3^2&=&{(r^2 + (\g_1+\g_2)^2\,\eta\,\cos^2\theta)\,(r^2 +
(\g_1+\g_2)^2\,\eta\,\sin^2\theta)\over 4}\,
\Bigl({dr^2\over r^2 + (\g_1+\g_2)^2\,\eta}+d\theta^2\Bigr)\nonumber\\
&+& {r^2\,( r^2 + (\g_1+\g_2)^2\,\eta)\over
4}\,\sin^2\theta\,\cos^2\theta\,d{\tilde\phi}^2
\eea

The metric
     $ds_3^2$ can be brought to the manifestly flat form
\be
ds_3^2=d{\tilde r}^2 + {\tilde r}^2 d{\tilde\theta}^2 + {\tilde r}^2
\sin^2{\tilde \theta} \,d{\tilde \phi}^2
\label{flat3}
\ee
by  the change of coordinates
\be
{\tilde r}={r^2 + (\g_1+\g_2)^2\,\eta\,\sin^2\theta \over 4}
\ee
\be{\tilde r}\,\cos^2{{\tilde \theta}\over 2}=
{r^2\over 4}\,\cos^2\theta
\ee

It is helpful to note several algebraic relations following from this
coordinate change. We have
\be
{\tilde r}\,\sin^2{{\tilde \theta}\over 2}=
{r^2+(\g_1+\g_2)^2\,\eta\over 4}\,\sin^2\theta
\ee
Defining the vector
\be
\vec{c}\equiv(0,0,c)\,,\quad c\equiv\Bigl({\g_1+\g_2\over
2}\Bigr)^{\!\!2}\,\eta
\ee
we get
\be
{\tilde r_c}\equiv |{\tilde x}+\vec{c}|={r^2 +
(\g_1+\g_2)^2\,\eta\,\cos^2\theta \over 4}
\ee
Then
\bea
&&{({\tilde r}+{\tilde r}_c-c) ({\tilde r}+{\tilde r}_c+c)\over
2\,{\tilde r}}=
{\tilde r}+{\tilde r}_c+c\,\cos{\tilde \theta}\nonumber\\
&&\cos{\tilde\theta}\,({\tilde r}+{\tilde r}_c)+c={{\tilde r}_c-{\tilde
r}\over c}\,
({\tilde r}+{\tilde r}_c+c\,\cos{\tilde \theta})
\eea
\be
f=4\,{{\tilde r}\,\g_1+{\tilde r}_c \,\g_2\over \g_1+\g_2}\,,\quad
\cos\,2\theta={{\tilde r}_c -{\tilde r}\over c}
\label{feq}
\ee

Using the coordinates ${\tilde r}$ and ${\tilde \theta}$
the Gibbons-Hawking potential $H_2$ and the magnetic potential $\chi$
become
\bea
H_2&=&{1\over \g_1+\g_2}\Bigl({\g_2\over{\tilde r}}+{\g_1\over {\tilde
r}_c}\Bigr)\nonumber\\
\chi&=&{\g_2\over \g_1+\g_2}\,\cos{\tilde\theta}+{\g_1\over
\g_1+\g_2}\,{{\tilde r}\,\cos{\tilde\theta}+c\over
{\tilde r}_c}
\label{h2}
\eea
In this form $H_2$ is explicitly harmonic.

It was shown in \cite{gmr} that the 6-D metric over a Gibbons-Hawking 
base can be expressed through a set of harmonic functions $H_i$ 
($i=1,\dots 6$) on $ds_3^2$. In the Appendix we find these harmonic 
functions $H_i$ for our geometries.

\section{The surface $f=0$}

\subsection{The `pseudo-hyperkahler' geometry of the base}

The field equations for  supersymmetric solutions  require the base to
be a 4-D hyperkahler manifold.
A conventional hyperkahler manifold has signature $(4,0)$; i.e.
$(++++)$. We can get a
similar geometry with signature
$(0,4)$ by simply reversing the sign of the 4-D metric. But consider eq.
(\ref{h2}) for $H_2$. From (\ref{sf}), (\ref{st}) we see that
$\gamma_1$ and $\gamma_2$ have opposite signs. This means that $H_2$
has opposite signs near $\tilde r, \tilde r_c$.
For $|\gamma_2|>|\gamma_1|$ we find that the change of sign happens over
a 2-D surface in the flat space
$ds_3^2$ in (\ref{gh}); this surface is topologically a $S^2$ which
surrounds the point $\vec c=(0,0,c)$.\footnote{This is easy to
see from the electrostatic analogy where we have charges $\gamma_2,
\gamma_1$ placed at $\tilde r, \tilde r_c$.}
For $|\gamma_1|>|\gamma_2|$ the $S^2$ surrounds the origin $(0,0,0)$.
    From  (\ref{gh}) we see that the base metric $h_{mn}dx^mdx^n$ has
signature
$(4,0)$ if we look at the region outside this $S^2$ and signature
$(0,4)$ inside. Since this is not the
nature of a traditional hyperkahler manifold we call the base
`pseudo-hyperkahler'.

    From the form (\ref{feq}) of $f$ we see that
\be
H_2={f\over 4 \tilde r \tilde r_c}
\ee
so that $H_2$ changes sign when $f$ passes through zero. Looking at the
complete metric (\ref{general6d})
and recalling that
\be
H=h=1+{Q\over f}
\ee
we see that the 6-D metric does not change signature; $H$ changes sign
at the same place that the base metric changes sign.\footnote{It was
shown in \cite{gms1,gms2} that $Q+f>0$ everywhere.}  This is of course
expected since the 6-D geometry has signature $(5,1)$ everywhere.

The condition $f=0$ will generically give  a 5-D surface in 6-D. The
place where the base metric  changes sign is
the intersection of the surface $f=0$ with the base. In the following we will
often describe the $f=0$ surface by giving the  4-D section obtained by restricting $f=0$ to  constant $t$.

In the 2-charge D1-D5 geometries the surface $f=0$ degenerates to a
closed curve $S^1$. This curve can be regarded as the center of a
`KK-monopole tube' (i.e. a KK monopole $\times ~S^1$) \cite{lmm}, and
different shapes of the $S^1$ give the set of different 2-charge
geometries \cite{lm4}. It thus appears interesting to investigate the
$f=0$ surface in more detail for the 3-charge case, which we do below.

\subsection{Topology of the surface $f=0$}

We have
\be
f=r^2+(\g_1+\g_2)\,\eta\,(\g_1\,\sin^2\theta+\g_2\,\cos^2\theta)
\ee
For generic values of $\g_1$ and $\g_2$, the equation $f=0$ defines a
hypersurface with no boundary. We are interested in the possible
significance of this surface in
terms of quantities in the dual CFT, so we take the limit $R>> \sqrt
{Q}$ which gives a large AdS type region and hence allows us to extract
the dual of the CFT
in the near horizon limit $r<<\sqrt{Q}$ \cite{lm6}. We see that in this
limit
\be
Q_p\ll Q, ~~~\eta\rightarrow 1
\ee

\subsubsection{The geometries (\ref{sf})}

Consider first the family (\ref{sf}).
In this near horizon  limit of (\ref{3chargemetric}) the change of
variables
\bea
&&\rho=k\,{r\over a}\,,\quad{\tilde t}={1\over k}\,{t\over R}\,,\quad
{\tilde y}={1\over k}\,{y\over R}\nonumber\\
&&{\tilde \psi}=\psi-{\g_1\over a}\, {t\over R}-{\g_2\over a} \,{y\over
R}=
\psi + n\, {t\over R}-
\Bigl(n+{1\over k}\Bigr)\, {y\over R}\nonumber\\
&&{\tilde \phi}=\psi-{\g_2\over a}\, {t\over R}-{\g_1\over a}\, {
y\over R}=\phi - \Bigl(n+{1\over k}\Bigl)\,
{t\over R} + n\, {y\over R}
\label{adsvar}
\eea
     gives the metric
\be
{ds^2_{n.h.}\over Q}= -(\rho^2+1)\,d{\tilde t}^2+{d
\rho^2\over
\rho^2+1}+\rho^2\,
d{\tilde y}^2 + d\theta^2 + \cos^2\theta\,d{\tilde \psi}^2 +
\sin^2\theta\,d{\tilde \phi}^2
\label{ads3}
\ee
which is locally but not globally equivalent to $AdS_3\times S^3$.
Indeed
the change of
variables (\ref{adsvar})  induces the following identifications
on $\tilde y$, ${\tilde \psi}$
and ${\tilde \phi}$:
\be
\Bigl({\tilde y\over R},{\tilde \psi},{\tilde \phi}\Bigr)\sim
\Bigl({\tilde y\over R},{\tilde
\psi},{\tilde \phi}\Bigr) + 2\pi m
\Bigl({1\over k},-{1\over k},0\Bigr),\quad m\in\mathbb{Z}
\label{adsorb}
\ee
and for $k>1$ the space (\ref{ads3}) is a $\mathbb{Z}_k$ orbifold of
$AdS_3\times
S^3$.

Consider now the hypersurface $f=0$ in this space. The topology of the
surface is different in the two cases (i) $n>0$, which gives $\g_2>0$,
$\g_1<0$  (ii) $n<0$, which gives $\g_2<0$ and $\g_1>0$.

     (i) The case $n>0$: Solving the equation
$f=0$ for $r$ in terms of $\theta$,
one sees that in the $(r,\theta)$ plane the surface is simply an
interval
\be
\theta\in I_{n>0}=[{\bar \theta},\pi/2]
\label{int1}
\ee
where
\be
{\bar \theta}=\mathrm{tan}^{-1} \Bigl(-{\g_2\over\g_1}\Bigr)^{1/2}=
\mathrm{tan}^{-1} \Bigl({k n+1 \over k n}\Bigr)^{1/2}
\ee
corresponds to $r=0$. Over this interval one has the three orthogonal
directions $\tilde y$, ${\tilde \psi}$
and ${\tilde \phi}$. The length of the circle ${\tilde \phi}$ never
vanishes over the interval (\ref{int1})
and thus it just gives  an overall $S^1$ factor. The cycle $\tilde y$
shrinks
at one end of the interval ($r=0$) and
${\tilde \psi}$ shrinks at the other end ($\theta=\pi/2$), so that they
form, together with the interval itself,
a sphere $S^3$.\footnote{This follows by comparison with the metric
$ds^2=d\theta^2+\cos^2\theta d\psi^2+\sin^2\theta
d\phi^2$ which gives $S^3$ for $0<\theta<{\pi\over 2}, 0<\psi<2\pi,
0<\phi<2\pi$. We easily check that the surface is smooth at the ends of
the interval (\ref{int1}) by inspection of the defining equation
$f=0$.} The identifications (\ref{adsorb}) act on this sphere
without fixed points since the $\tilde y$ and $\tilde \psi$ circles do 
not shrink at the same time. Thus
the resulting orbifold space is the (smooth) Lens space
$S^3/\mathbb{Z}_k$. We conclude that in the case $n>0$
the hypersurface $f=0$ is topologically $S^1\times (S^3/\mathbb{Z}_k)$.

(ii) The case $n<0$: The interval in the $(r,\theta)$ plane is
\be
\theta\in I_{n<0}=[0,{\bar \theta}]
\label{int2}
\ee
This time the $\tilde\psi$ circle remains of nonzero size everywhere 
and provides an $S^1$ factor.
At  $\theta=0$  it is the ${\tilde \phi}$ circle that
shrinks, and at $\theta=\bar\theta$ the $\tilde y$ circle shrinks.
  We thus get
$S^1\times S^3$, with the $S^1$ spanned by $\tilde \psi$ and
the
$S^3$  made up by $\tilde y$,
$\tilde \phi$ and the interval $I_{n<0}$. This surface is to be divided 
by the orbifold action
(\ref{adsorb}) which this time acts on both  the
$S^1$ and the $S^3$. Noting that the $\tilde\psi$ circle does not 
shrink anywhere we see that
the orbifold action has no fixed points, and the resulting hypersurface 
is
a smooth orbifold
$(S^1\times S^3)/\mathbb{Z}_k$.

\subsubsection{The geometries (\ref{st})}

A similar analysis can be done for the $f=0$ surface for the family of
geometries with $\gamma_1, \gamma_2$ given by (\ref{st}). In the near
horizon limit these geometries
reduce again to the $AdS_3\times S^3$ form (\ref{ads3})
after the  coordinate transformation
\bea
&&\rho=n_5\,n\,(k n+1)\,{r\over a}\,,\quad{\tilde t}={1\over n_5\,n\,(k
n+1)}\,{t\over R}\,,\quad
{\tilde y}={1\over n_5\,n\,(k n+1)}\,{y\over R}\nonumber\\
&&{\tilde \psi}=
\psi + {k\over n_5\,(k n+1)}\, {t\over R}-
{1\over n_5\,n}\, {y\over R}\,,\quad{\tilde \phi}=\phi - {1\over
n_5\,n}\,
{t\over R} + {k\over n_5\,(k n+1)}\, {y\over R}
\label{adsvarst}
\eea
We see from  (\ref{adsvarst}) that these coordinates are subject to the
identifications
\be
\Bigl({\tilde y\over R},{\tilde \psi},{\tilde \phi}\Bigr)\sim
\Bigl({\tilde y\over R},{\tilde
\psi},{\tilde \phi}\Bigr) + 2\pi m
\Bigl({1\over n_5\,n\,(k n+1)},-{1\over n_5\,n},{k\over n_5\,(k
n+1)}\Bigr),\quad m\in\mathbb{Z}
\label{adsorbst}
\ee
We thus get a $\mathbb{Z}_{n_5 n\,(k n+1)}$ orbifold group; let
$\omega$ be the generator of this group.  We examine  the
cases $n>0$ and
$n<0$ separately.

(i) The case $n>0$:
The intersection of the surface $f=0$ with the $(r,\theta)$ plane is
again
the interval $I_{n>0}$ defined in (\ref{int1}). As in the previous
case, the circles
$\tilde y$ and $\tilde \psi$  fibered over $I_{n>0}$ form an $S^3$ and
$\tilde \phi$ gives
a finite size $S^1$. However the orbifold group $\mathbb{Z}_{n_5 n\,(k
n+1)}$ now acts
simultaneously on the $S^3$ and the $S^1$, according to
(\ref{adsorbst}),
and the $f=0$ surface is an orbifold $(S^3\times S^1)/\mathbb{Z}_{n_5
n\,(k n+1)}$ with orbifold action given by
(\ref{adsorbst}).

This orbifold action again has no fixed points. To see this, consider
first the place where the $\tilde y$ circle shrinks
to zero. To get a fixed point we would need to get a trivial action on
the two nonshrinking circles $\tilde \psi, \tilde \phi$.
This means that we must take an element of the orbifold group
$\omega^s$, where ${s\over n_5 n}$ and ${k s\over n_5(kn+1)}$ are both
integers. A little thought then shows that  ${s\over n_5 n (kn+1)}$
must be integral as
well, so that we get only the trivial element of the orbifold group. A
similar analysis shows that there is no fixed point when the $\tilde
\psi$ circle shrinks. Thus there is no fixed point of the orbifold
group and the surface $f=0$ is smooth.

(ii) The case $n<0$: As in the previous case, the $f=0$ surface is an
orbifold
$(S^3\times S^1)/\mathbb{Z}_{n_5 n\,(k n+1)}$, but this time the $S^3$
is made up by
the circles $\tilde y$ and $\tilde \phi$ fibered over the interval
$I_{n<0}$ of Eq. (\ref{int2})
and the $S^1$ is generated by $\tilde \psi$. The orbifold action is 
given by (\ref{adsorbst}).
An analysis similar to the one for case (i) shows that there are no fixed 
points of the
orbifold action and the surface $f=0$ is smooth.

\section{Regularity of the 3-charge solution}

It was found in \cite{gms1, gms2} that the metric (\ref{3chargemetric})
is regular everywhere apart from possible  orbifold singularities. We
wish to examine this singularity structure from the point of view of
the base-fiber decomposition
(\ref{general6d}) of the metric; this might help us to understand how
to find more general bound states using a formalism like \cite{gmr}.

\subsection{The geometries (\ref{sf})}

We first look at the geometries with $\gamma_1, \gamma_2$ given
by  (\ref{sf}). We know from \cite{gms2} that these  are completely
smooth spaces for $k=1$ while for
$k>1$ they have an orbifold singularity along an $S^1$. (Transverse to
the $S^1$ this is an ALE type singularity which arise from the
collision of $k$ KK monopoles \cite{lmm}.)

Let us first start by looking at the base metric. As can be seen from
the form of the Gibbons-Hawking potential $H_2$
in (\ref{h2}), potential singularities of the base can occur
at:
i) $\tilde r=0$ (i.e. $r=0$ and $\theta=0$); ii) ${\tilde r}_c=0$
(i.e. $r=0$ and $\theta=\pi/2$); iii) $\g_2\,{\tilde r}_c+\g_1\,{\tilde
r}=0$ (i.e. $f=0$). We will  examine
cases (i) and (ii) here and return to (iii) in subsection
(\ref{fto0limit}) below.

\subsubsection{The ${\tilde r}\to 0$ limit}

We find that the base space metric has in general an orbifold
singularity at $\tilde r=0$. This is seen as follows.
Around the point $\tilde r=0$  define the new coordinates $r'$,
$\theta'$, $\psi'$ and $\phi'$ as follows\footnote{For definiteness we
are assuming here that $n>0$ and thus $\g_2>0$ and $\g_1<0$. In the
case $n<0$ the
appropriate definition of $r'^2$ has a negative sign compared to the
definition in (\ref{basecoord}) and the base
space has signature $(0,4)$ around ${\tilde r}=0$.}
\bea
&& r'\,^2 = 4\,{\g_2\over \g_1+\g_2}\,{\tilde r}= 4\,(k n +1)\,{\tilde
r}\,,\quad
\theta'={\tilde\theta\over 2}\nonumber\\
&&\psi'={\g_1+\g_2\over \g_2}\,\psi={1\over k n +1}\,\psi \,,\quad
\phi'= -{\g_1\over \g_2}\,\psi + \phi=
{k n\over k n+1}\,\psi+\phi
\label{basecoord}
\eea
In these coordinates the base metric is
\be
h_{mn} dx^m dx^n\approx dr'\,^2 +
r'\,^2\,d\theta'\,^2+r'\,^2\,\cos^2\theta'\,d\psi'\,^2+r'\,^2\,\sin^2\theta'
\,d\phi'\,^2
\ee
which is the form of the metric for flat $\mathbb{R}^4$. However the
new angular coordinates $\psi'$ and $\phi'$ are subject
to the  identifications
\be
(\psi',\phi')\sim (\psi',\phi')+2\pi n_1 \Bigl({1\over k n+1},1-{1\over
k n+1}\Bigr)+2\pi n_2(0,1),\quad n_1,n_2
\in\mathbb{Z}
\label{orbifold}
\ee
so that the base metric around ${\tilde r}=0$ is equivalent to an
orbifold $\mathbb{R}^4/\mathbb{Z}_{k n+1}$, where
the group $\mathbb{Z}_{k n+1}$ acts on both the $\psi'$ and $\phi'$
cycles according to the identifications
(\ref{orbifold}). Since at ${\tilde r}=0$ both these cycles
shrink to a point, ${\tilde r}=0$ is a fixed point for the orbifold
action and the space has an ALE type singularity.

In the total 6-D space however there is no singularity at  $\tilde
r=0$. To see this, consider the  behavior of $H$, $F$, $\beta$ and
$\omega$  around $\tilde r=0$
\bea
&&H\approx 1+{Q\over \g_2\,(\g_1+\g_2)\,\eta}\,,\quad {F\over 2}\approx
{\g_1\over (\g_1+\g_2)\,\eta}\nonumber\\
&&\sqrt{2}\,\beta\approx -\sqrt{2}\,\omega \approx {Q\over
\g_2}\,d\psi\equiv \sqrt{2}\,\beta_\psi\,d\psi
\label{rto0}
\eea
The 6-D metric thus has the form
\bea
&&ds^2\approx \Bigl(1+{k^2 Q\over a^2\eta\,(k n+1)}\Bigr)^{-1}
\,\Bigl[(-dt^2+dy'\,^2)+{k n\over \eta}\,
(dt-dy')^2\Bigr]\\
&&\qquad+\Bigl(1+{k^2 Q\over a^2\eta\,(k n+1)}\Bigr)\,[dr'\,^2 +
r'\,^2\,d\theta'\,^2+r'\,^2\,\cos^2\theta'\,d\psi'\,^2+
r'\,^2\,\sin^2\theta'\,d\phi'\,^2]\nonumber
\label{totalrto0}
\eea
where we have defined
\be
{y'\over R}= {y\over R} - {\sqrt{2}\,\beta_{\psi}\over R}\,\psi =
{y\over R}-{k\over k n+1}\,\psi
\label{yprime}
\ee
Note that in these variables the $t, y'$ subspace is orthogonal to the
other four coordinate directions.  These
variables are subject to the identifications
\be
\Bigl(\psi',\phi',{y'\over R}\Bigr)\sim \Bigl(\psi',\phi',{y'\over
R}\Bigr)+2\pi n_1
\Bigl({1\over k n+1},1-{1\over k n+1},-{k\over k n+1}\Bigr)
\ee
The $y'$ circle does not shrink to a point  at $\tilde r=0$. Since $k$
and $k n + 1$ cannot share a common factor, we see that there will be a
nonzero shift in the $y'$ direction when we move from one orbifold
image of $(\psi', \phi')$ to another.
Thus $\mathbb{Z}_{k n+1}$ does not have fixed points on the total
space. We conclude that the 6-dimensional metric
(\ref{3chargemetric}) is regular around ${\tilde r}=0$.

\subsubsection{The ${\tilde r}_c\to 0$ limit}

We similarly find an orbifold singularity in the base space metric at
$\tilde r_c=0$, but in this case we will also encounter an orbifold
singularity in the total 6-D space if $k>1$. This latter singularity is
     just the orbifold singularity found for the total space in
\cite{gms2}.

This analysis around $\tilde r_c=0$  is connected to the analysis
around $\tilde r=0$ by exchanging $\g_1$ with $\g_2$, $\psi$ with 
$\phi$,
$\theta$ with $\pi/2-\theta$.
At the point ${\tilde r}_c=0$ the base metric is reduced to locally
flat form -- but in this case
with signature $(0,4)$ -- by the change of coordinates
\bea
&& r''\,^2 = -4\,{\g_1\over \g_1+\g_2}\,{\tilde r}_c= 4\,k n \,{\tilde
r}\,,\quad
\theta''={{\tilde\theta}_c\over 2}\nonumber\\
&&\psi''=\psi -{\g_2\over \g_1}\,\phi=\psi + {k n + 1\over k n}\,\phi
\,,\quad \phi''=
{\g_1+\g_2\over \g_1}\,\phi=- {1\over k n}\,\phi
\label{basecoordrc}
\eea
where ${\tilde\theta}_c$ is the polar coordinate around the point
${\tilde r}_c=0$:
\be
\cos\,{\tilde \theta}_c = {{\tilde r}\,\cos\,\tilde\theta +c\over
{\tilde r}_c}
\ee
The periodic identifications on the $\psi''$ and $\phi''$ coordinates
\be
(\psi'',\phi'')\sim (\psi'',\phi'')+2\pi n_1 (1,0)+ 2\pi n_2
\Bigl(1+{1\over k n},-{1\over k n}\Bigr), \quad n_1,n_2
\in\mathbb{Z}
\ee
give the orbifold group $\mathbb{Z}_{k n}$. The group action has a
fixed point
at ${\tilde r}_c=0$ and the base space has an ALE singularity at this
point.

Let us now look at the total 6-D space. Around ${\tilde r}_c=0$ we have
\bea
&&H\approx 1+{Q\over \g_1\,(\g_1+\g_2)\,\eta}\,,\quad {F\over 2}\approx
{\g_2\over (\g_1+\g_2)\,\eta}\nonumber\\
&&\sqrt{2}\,\beta\approx -\sqrt{2}\,\omega \approx {Q\over
\g_1}\,d\phi\equiv \sqrt{2}\,\beta_\phi\,d\phi
\label{rcto0}
\eea
and thus the metric of the total space is\footnote{Note that ${k
Q\over a^2\eta\,n}-1={1\over na^2}[kQ+na^2(2kn+1)]$
which is positive since we are working with $n>0$.  The signature of
the total space is thus everywhere $(5,1)$.}
\bea
&&ds^2\approx \Bigl({k Q\over a^2\eta\,n}-1\Bigr)^{-1}
\,\Bigl[(dt^2-dy''\,^2)+{k n +1 \over \eta}\,(dt-dy'')^2\Bigr]
\\
&&\qquad+\Bigl({k Q\over a^2\eta\,n }-1\Bigr)\,[dr''\,^2 +
r''\,^2\,d\theta''\,^2+r''\,^2\,\cos^2\theta''\,d\psi''\,^2+
r''\,^2\,\sin^2\theta''\,d\phi''\,^2]\nonumber
\label{totalrcto0}
\eea
where
\be
{y''\over R}= {y\over R} - {\sqrt{2}\,\beta_{\phi}\over R}\,\phi =
{y\over R}+{1\over n}\,\phi
\label{ydoubleprime}
\ee
Thus the 6-dimensional space defined by the metric (\ref{totalrcto0})
is the quotient
of a smooth space by the $\mathbb{Z}_{k n}$ group that identifies the
points
\be
\Bigl(\psi'',\phi'',{y''\over R}\Bigr)\sim
\Bigl(\psi'',\phi'',{y''\over R}\Bigr)+2\pi n_2
\Bigl(1+{1\over k n},-{1\over k n}, {1\over n}\Bigr)
\ee
Let $\omega$ be the generator of this orbifold group. The element 
$\omega^n$ acts trivially on $y''$.
The fixed point group is thus $\mathbb{Z}_{k }$, and we have  an 
orbifold singularity in the 6-D space of order $k$ if $k>1$.

\subsection{The geometries (\ref{st})}

We can do a similar analysis for the geometries with $\gamma_1,
\gamma_2$ given by
(\ref{st}). We will again find that there are orbifold singularities in the
base metric at both
$\tilde r=0$ and $\tilde r_c=0$ (in fact the orbifold groups are the
same as those for the geometries
(\ref{sf})). But the full 6-D metric also has orbifold singularities at
each of these locations, while the class
(\ref{sf})  had an orbifold singularities only at $\tilde r_c=0$. These
singularities of the full 6-D metric are of course just the same ones
found directly in \cite{gms2}, but here we are interested in seeing how
they arise from the base $\times$ fiber decomposition of the metric.

Consider first the location ${\tilde r}=0$. We insert the values
of $\g_1,
\g_2$ given in (\ref{st})  into the expressions (\ref{basecoord},
\ref{rto0}), and find  the following form of the 6-D metric near
$\tilde r=0$
\bea
&&ds^2\approx \Bigl(1+{n^2\, n_5^2 \,(k n+1) Q\over a^2\eta}\Bigr)^{-1}
\,
\Bigl[(-dt^2+dy'\,^2)+{k n\over \eta}\,
(dt-dy')^2\Bigr]\\
&&\qquad+\Bigl(1+{n^2\,n_5^2\,(k n+1) Q\over a^2\eta}\Bigr)\,
[dr'\,^2 + r'\,^2\,d\theta'\,^2+r'\,^2\,\cos^2\theta'\,d\psi'\,^2+
r'\,^2\,\sin^2\theta'\,d\phi'\,^2]\nonumber
\label{totalrto0st}
\eea
The variables here are now defined as
\bea
&&r'\,^2 = 4\,(k n +1)\,{\tilde r}\,,\quad
\theta'={\tilde\theta\over 2}\nonumber\\
&&\psi'={1\over k n+1}\,\psi\,,\quad \phi'={k n\over k
n+1}\,\psi+\phi\,,\quad
{y'\over R}={y\over R}- n_5\,n\,\psi
\eea
and are thus subject to the identifications
\be
\Bigl(\psi',\phi',{y'\over R}\Bigr)\sim \Bigl(\psi',\phi',{y'\over
R}\Bigr)+
2\pi n_1
\Bigl({1\over k n+1},1-{1\over k n+1},-n_5\,n\Bigr)
\ee
If we look at just the coordinates $\psi', \phi'$ which lie in the base
metric, then since both these circles shrink to zero at $\tilde r=0$  we
find an orbifold singularity $\mathbb{Z}_{kn+1}$. This is the same
singularity that we found for the base when looking at the family
(\ref{sf}).\footnote{The definitions $\psi', \phi'$ in
(\ref{basecoord}) involve the {\it ratios} of the $\gamma_i$, and these
ratios are the same for the families (\ref{sf}) and (\ref{st}); thus
the singularities of the base will turn out to be the same in the two
cases.} Since $y'$ shifts by an integer multiple of $2\pi$ under this orbifold action, 
the full 6-D space also has an orbifold singularity $\mathbb{Z}_{kn+1}$
at $\tilde r=0$.

Let us now examine the singularities at $\tilde r_c=0$.
From
(\ref{basecoordrc}, \ref{rcto0}), we find that the
the metrics (\ref{st}) behave as
\bea
&&\!\!\!\!ds^2\approx \Bigl({n\, n_5^2 \,(k n+1)^2 Q\over a^2
k\eta}-1\Bigr)^{-1}
\,\Bigl[(dt^2-dy''\,^2)+{k n +1 \over \eta}\,
(dt-dy'')^2\Bigr]\\
&&\!\!\!\!\qquad+\Bigl({n\, n_5^2 \,(k n+1)^2 Q\over a^2 k\eta}-1\Bigr)\,
[dr''\,^2 + r''\,^2\,d\theta''\,^2+r''\,^2\,\cos^2\theta''\,d\psi''\,^2+
r''\,^2\,\sin^2\theta''\,d\phi''\,^2]\nonumber
\label{totalrcto0st}
\eea
where
\bea
&& r''\,^2 = 4\,k n \,{\tilde r}\,,\quad
\theta''={{\tilde\theta}_c\over 2}\nonumber\\
&&\psi''=\psi + {k n + 1\over k n}\,\phi \,,\quad \phi''=- {1\over k
n}\,\phi\,,\quad {y''\over R}={y\over R}+{n_5\,(k n+1)\over k}\,\phi
\eea
These coordinates are subject to the identifications
\be
\Bigl(\psi'',\phi'',{y''\over R}\Bigr)\sim \Bigl(\psi'',\phi'',{y''\over
R}\Bigr)+2\pi n_2
\Bigl(1+{1\over k n},-{1\over k n}, {n_5 (k n+1)\over k}\Bigr)
\label{action1}
\ee
The base metric has an orbifold singularity $\mathbb{Z}_{k n}$ since
the $\psi'', \phi''$ circles shrink at ${\tilde r}_c=0$. The $y''$
coordinate shifts under the orbifold action however, and the $y''$
circle does not shrink here. Let $\omega$ be the generator of the
orbifold action (\ref{action1}). If $n_5$, $k$ share no common factor
then points of the full 6-D metric are left invariant by $\omega^k$,
and we find an orbifold singularity $\mathbb{Z}_{ n}$. (There is of
course no singularity if $n=1$.) If $n_5$ and $k$ share a common factor
$m$
then $y''$ is left invariant under $\omega^{k\over m}$, and we have an
orbifold singularity $\mathbb{Z}_{nm}$.

\subsection{The $f\to 0$ limit}
\label{fto0limit}

We have observed that the Gibbons-Hawking base space
  is such that at the intersection with the hypersurface $f=0$
the Gibbons-Hawking
     potential $H_2$ vanishes and the base metric has a  severe
singularity: its signature changes from $(4,0)$ to $(0,4)$.
   From the analysis in \cite{gms1, gms2} we know that the 6-D metric
(\ref{3chargemetric}) has no singularity at $f=0$, so  singularities of
the base must cancel contributions from the fiber in some way to make
the total space smooth.
We study the cancellation of the leading order terms in some detail,
since we expect that there will be an analogue of the surface $f=0$
for more general 3-charge geometries, and the fiber and base should to
be such that the singularities cancel.

Note that $H$, $F$ and $\beta$ have simple poles at $f=0$,
$\omega$ has a double pole proportional to $Q_p$, while $H_2$ vanishes 
linearly.
\bea
&&H\to {Q\over f}\,,\quad {F\over 2}=-{Q_p\over f}\nonumber\\
&&\sqrt{2}\,\beta={Q\over f}\,(\g_1 + \g_2)\,\eta\,(\cos^2\theta\,d\psi
+
\sin^2\theta\,d\phi)\nonumber\\
&&\sqrt{2}\,\omega\to {2 Q\,Q_p\over f^2}\,(\g_1 + \g_2)\,\eta\,
(\cos^2\theta\,d\psi + \sin^2\theta\,d\phi)\nonumber \\
&&H_2\to-{\gamma_2 \over 4\tilde r^2\, \gamma_1} f
\label{fto0}
\eea
We observe that in this limit the 1-forms  $\beta$ and $\omega$ are
parallel, being proportional to
$\cos^2\theta\,d\psi + \sin^2\theta\,d\phi$. The part of the base
metric which diverges is also
parallel to this form, a fact which can be seen as follows. As $f\to 0$
\be
{\tilde r\over {\tilde r}_c}\to -{\g_2\over \g_1}\,,\quad
\chi\to-{(\g_1+\g_2)\g_2\eta\over 4\tilde r}\,,\quad \cos 2\theta\to
-{4\tilde r\over (\g_1+\g_2)\g_2\eta}
\ee
and thus
\be
\cos^2\theta\,d\psi+\sin^2\theta\,d\phi=
{d\tilde\phi+\cos 2\theta\,d\tau\over 2}\to -
{2\tilde r\over (\g_1+\g_2)\g_2\eta}\,(d\tau + \chi\,d{\tilde \phi})
\ee
and we note that the part $(d\tau + \chi\,d{\tilde \phi})^2$ of the
base metric is the one with the divergent coefficient.

We  rewrite (\ref{fto0})  as
\bea
\sqrt{2}\,\beta&\to&-{2 Q\,\tilde r\over \g_2\,f}\,(d\tau +
\chi\,d{\tilde
\phi})\equiv
\sqrt{2}\,\beta_0\,(d\tau + \chi\,d{\tilde \phi})\nonumber\\
\sqrt{2}\,\omega&\to&-{4 Q\,Q_p\,\tilde r\over \g_2\,f^2}\,(d\tau +
\chi\,d{\tilde \phi})\equiv
\sqrt{2}\,\omega_0\,(d\tau + \chi\,d{\tilde \phi})
\eea

We can now substitute the above expressions into (\ref{general6d}) and
collect the coefficients of the leading divergence as $f\rightarrow 0$.
  We get
\be
ds_{6D}^2\rightarrow
[-H^{-1}\,\sqrt{2}\,\beta_0\,\Bigl(\sqrt{2}\,\omega_0 + {F\over
2}\,\sqrt{2}\,\beta_0\Bigr)+H\,H_2^{-1}]\,(d\tau + \chi\,d{\tilde
\phi})^2
\ee
While each term in the above expression diverges as $\sim {1\over
f^2}$, we find using the above
limiting forms of the coefficient functions that the coefficient of
${1\over f^2}$ cancels. Note that this cancellation occurs
for generic values of $\g_1$ and $\g_2$.

Solutions with Gibbons-Hawking base can be written in terms of harmonic 
functions $H_i$ on $ds_3^2$ \cite{gmr}
(the $H_i$ for our bound states are given in the Appendix). If we try 
to make new geometries by
superposing the harmonic functions found for bound states then we do 
not in general get a cancellation of this
${1\over f^2}$ singularity. Thus we cannot easily make new bound state 
geometries by exploiting the  linearity
of the space of harmonic functions.

\section{Discussion}

It is interesting that supersymmetric solutions in 6-D (with
translation symmetry along $t,y$) can be written as
a hyperkahler base times a $(t,y)$ fiber. Further, the equations of
motion allow the coefficient functions in this
decomposition to be expressed in terms of a set of harmonic functions
\cite{gmr,bw}.

Large classes of classical 3-charge extremal solutions may be written
down using such formalisms \cite{bw,emparan3,gg}, but to address
the questions relevant to black holes we need to find the solutions
that correspond to actual {\it bound} states
of the D1,D5,P charges. A generic choice of harmonic functions would
not give a bound state; we can already see this from the 2-charge D1-D5
case where we can superpose harmonic functions to make non-bound states
from bound states.
In particular (as noted in the Appendix)  superposing harmonic
functions to create new  3-charge geometries gives solutions that have
pathologies and are not likely to be bound states.

To gain some insight into the nature of the bound state geometries we
have in this paper expressed the geometries of \cite{gms1, gms2} in the
`base $\times$ fiber' form. These geometries are known to represent
bound states since they were obtained by applying exact symmetries
(spectral flow and S,T dualities) to 2-charge bound state geometries.
(These 2-charge geometries in turn were known to be bound states
because  they were generated by S,T dualities applied
to a single fundamental string carrying vibrations \cite{lm4}.)

   From this analysis we observed that the base is not hyperkahler but
`pseudo-hyper\-kahler'; the signature
changes from $(4,0)$ to $(0,4)$ across a hypersurface in the base. This
hypersurface is the intersection of the base with the surface $f=0$
defined in the full 6-D space. In the 2-charge case the surface $f=0$
(when restricted to $t=const.$)
degenerates to a simple closed curve $S^1$, and different shapes of
this curve map out the different microstates \cite{lm4}. It is thus
interesting to investigate the $f=0$ surface in the 3-charge case as
well. We found that in the 3-charge case
the $f=0$ surface (again restricted to $t=const.$) was $S^1\times S^3$
divided by an orbifold group. This group acted without fixed points, so
the $f=0$ surface was smooth. Roughly speaking, the $S^1$ is the same
$S^1$ as in the 2-charge case, and the $S^3$ collapses to a point when
the momentum charge vanishes. It is possible that different shapes of
the $f=0$ surface give the different 3-charge microstates, just as
different shapes of the $f=0$ curve in the 2-charge case described the
microstates.

It is interesting to compute the area of the $f=0$ hypersurface (at 
fixed $t$).
For the family (\ref{sf}) we find $A=4\pi^2(2\pi R){\sqrt{kn+1}-1\over 
2kn+1}\sqrt{Q_1Q_5Q_p}$
(we have replaced $Q^2$ by $Q_1Q_5$ to achieve a form more symmetric 
between the charges).
Dividing by the 6-D Newton constant we get
\be
{A\over G^{(6)}}\sim {\sqrt{Q_1Q_5Q_p}\over G^{(5)}}
\ee
where $G^{(5)}=G^{(6)}/(2\pi R)$ is the 5-D Newton constant obtained 
upon reduction along $y$.
The above relation is reminiscent of the  entropy relation for the 
black hole in 5-D, though we do not have
any horizon in the geometry. It will be interesting to see if a surface 
analogous to $f=0$ with such an area exists for
generic 3-charge microstates, and to relate this surface to the 
supertube description of bound states \cite{supertubes,bk,bena,marolf,bak}.

To gain further insight into the base $\times$ fiber form, we analyzed
the singularities occurring in the base metric. The base is in our case
a Gibbons-Hawking geometry, with the harmonic function having two
centers. The full 6-D geometry is known to have orbifold singularities
except in special cases. In the base $\times $ fiber decomposition we
find that there are orbifold singularities in the base at the two
centers of the harmonic function. We saw explicitly how putting
together the base with the fiber generates the appropriate orbifold
singularities for the full 6-D metric. If one could understand the 
general singularity structure of the base
and the regularity conditions needed on the full 6-D metric then one 
might be able to extract
the D1-D5-P bound state geometries out of  the general family of 
supersymmetric solutions.

\section*{Acknowledgments}

S.G. was supported by  an I.N.F.N. fellowship. The work of S.D.M  was
supported in part by DOE grant DE-FG02-91ER-40690. We
thank Ashish Saxena, Yogesh Srivastava and Nick Warner for many helpful
discussions.

\appendix
\section{Harmonic functions for the 3-charge metric}
\renewcommand{\theequation}{A.\arabic{equation}}
\setcounter{equation}{0}

In \cite{gmr} an interesting observation was made: When the base metric
is of Gibbons-Hawking type, as  turns out to be the
case for our 3-charge metric (\ref{3chargemetric}), one can rewrite the
quantities $H$, $F$, $\beta$, $\omega$ appearing in
(\ref{3chargegmr}) as some combinations of harmonic functions on flat
$\mathbb{R}^3$. Conversely,  any choice of these
harmonic functions generates a solution that will satisfy the
equations of motion
(\ref{beta}--\ref{f}).

In this appendix, we will derive the harmonic functions corresponding
to the 3-charge metric (\ref{3chargegmr}).

Let us introduce a convenient basis of 1-forms
\be
\sigma\equiv d\tau + \chi\,d{\tilde\phi}\,,\quad d{\tilde
\alpha}^i\equiv\{d{\tilde r},d{\tilde\theta},d{\tilde\phi}\}
\ee
in terms of which the $\star$ operation reads (only the components
which are used in the following computation
are written down)
\bea
&&\star \,(d{\tilde r}\,\sigma) = -H_2\,{\tilde r}^2\,\sin\,{\tilde
\theta}\,d{\tilde\theta}\,d{\tilde \phi}\,,\quad
     \star \,(d{\tilde \theta}\,\sigma) = H_2\,\sin\,{\tilde
\theta}\,d{\tilde r}\,d{\tilde \phi}\nonumber\\
&&\star \,(d{\tilde r}\,d{\tilde\phi}) = H_2^{-1}\,\sin^{-1}{\tilde
\theta}\,d{\tilde\theta}\,\sigma\,,\quad
     \star \,(d{\tilde \theta}\,d{\tilde\phi}) =
-H_2^{-1}\,r^{-2}\,\sin^{-1}{\tilde \theta}\,d{\tilde r}\,\sigma
\eea

Let us expand the 1-forms $\beta$ and $\omega$ and the self-dual 2-form
$\mathcal{G}^+$ in this basis\footnote{The 3-dimensional indices $i$,
$j$, $k$
are raised and lowered with the 3-dimensional flat metric $g_3$ defined
in (\ref{flat3}), and
$\epsilon_{{\tilde r}{\tilde \theta}{\tilde \phi}}
=\sqrt{g_3}={\tilde r}^2\sin\tilde\theta$.}
\be
\beta=\beta_0\,\sigma+\beta_i\,d{\tilde \alpha}^i\,,\quad
\omega=\omega_0\,\sigma+\omega_i\,d{\tilde \alpha}^i
\ee
\be
\mathcal{G}^+=
C_i\,d{\tilde \alpha}^i\,\sigma - {\epsilon_{ij}^{~~k}\over 2}\,C_k\,
H_2\,d{\tilde \alpha}^i\,d{\tilde \alpha}^j
\ee
where we used self-duality of $\mathcal{G}^+$.

As shown in \cite{gmr}, the equation of motion (\ref{beta}) implies
\be
\beta_0=H_2^{-1}\,H_3\,,\quad \star_3 d(\beta_i d{\tilde \alpha}^i) =
-dH_3
\label{beta2}
\ee
with $H_3$ harmonic on $\mathbb{R}^3$. For the 3-charge metric
(\ref{3chargemetric}) we have
\be
\sqrt{2}\,\beta_0={Q\,(\g_1+\g_2)\,\eta\over 2}\,{\cos\,2\theta\over f}=
{Q\over 2}\,{{\tilde r}_c-{\tilde r}\over \g_1\,{\tilde
r}+\g_2\,{\tilde r}_c}
\ee
from which we find
\be
\sqrt{2}\,H_3 = {Q\over 2(\g_1+\g_2)}\,\Bigl({1\over{\tilde
r}}-{1\over{\tilde r}_c}\Bigr)
\ee

The closure of $\mathcal{G}^+$, Eq.(\ref{g}), implies
\be
C_i=\partial_i(H_2^{-1}\,H_4)
\ee
with $H_4$ harmonic on $\mathbb{R}^3$.
The value of $\mathcal{G}^+$ for the metric (\ref{3chargemetric}) can be
computed from the definition (\ref{g+}) and from (\ref{3chargegmr}).
The result for $C_i$, derived
with the help of Mathematica, is
\bea
&&\sqrt{2}\,C_{\tilde r}={(\g_1+\g_2)^3\,\g_1\g_2\,\eta\over
16}\,{4{\tilde r}\,\cos\,{\tilde\theta}+(\g_1+\g_2)^2\,\eta
\over {\tilde r}_c\,(\g_1\,{\tilde r}+\g_2\,{\tilde r}_c)^2}\nonumber\\
&&\sqrt{2}\,C_{\tilde \theta}={(\g_1+\g_2)^3\,\g_1\g_2\,\eta\over
4}\,{{\tilde r}^2\,\sin\,{\tilde\theta}
\over {\tilde r}_c\,(\g_1\,{\tilde r}+\g_2\,{\tilde r}_c)^2}\nonumber\\
&&\sqrt{2}\,C_{\tilde\phi}=0
\eea
   From this we see that
\be
\sqrt{2}\,C_i=\partial_i\,\Bigl[{\g_1+\g_2\over 2}\,{\g_1\,{\tilde
r}-\g_2\,{\tilde r}_c\over
\g_1\,{\tilde r}+\g_2\,{\tilde r}_c}\Bigr]
\ee
and thus
\be
\sqrt{2}\,H_4={1\over2}\,\Bigl({\g_1\over{\tilde
r}_c}-{\g_2\over{\tilde r}}\Bigr)+\mathrm{const.}\,H_2
\ee
The value of the constant in the equation above can be
chosen in such a way that $H_4$ vanishes in the two charge
limit ($\g_1\g_2=0$):
\be
\mathrm{const.}={\g_2-\g_1\over 2}
\ee
With this choice we get
\be
\sqrt{2}\,H_4={\g_1\g_2\over \g_1+\g_2}\,\Bigl({1\over{\tilde
r}_c}-{1\over{\tilde r}}\Bigr)
\ee

The remaining equations of motion (\ref{h}) and (\ref{f}) imply
\be
H=H_1+H_2^{-1}\,H_3\,H_4
\ee
\be
F=-H_5-H_2^{-1}\,H_4^2
\ee
where again $H_1$ and $H_5$ are harmonic on $\mathbb{R}^3$. The
values for the our
3-charge metrics are
\be
H=1+{Q\over 4}\,{\g_1+\g_2\over\g_1\,{\tilde r}+\g_2\,{\tilde
r}_c}\,\Longrightarrow\, H_1=1+{Q\over 4(\g_1+\g_2)}\,
\Bigl({\g_1\over{\tilde r}}+{\g_2\over{\tilde r}_c}\Bigr)
\ee
\be
{F\over 2}={\g_1\g_2\over 4}\,{\g_1+\g_2\over\g_1\,{\tilde
r}+\g_2\,{\tilde r}_c}\,\Longrightarrow\,
     H_5=-{\g_1\g_2\over 2(\g_1+\g_2)}\,
\Bigl({\g_1\over{\tilde r}}+{\g_2\over{\tilde r}_c}\Bigr)
\ee

Finally, from the definition (\ref{g+}) of $\mathcal{G}^+$ we find that
\bea
\omega_0&=&H_6 + H_2^{-2}\,H_3\,H_4^2+ H_2^{-1}\,H_1\,H_4 + {1\over 2}
H_2^{-1}\,H_3\,H_5\\
\star_3 d(\omega_i d{\tilde
\alpha}^i)&=&H_2\,d\omega_0-\omega_0\,dH_2-2(H_1
H_2+ H_3 H_4)\,d\Bigl[{H_4\over H_2}\Bigr]-
(H_4^2+H_2 H_5)\,d\Bigl[{H_3\over H_2}\Bigr]\nonumber
\label{omega2}
\eea
where $H_6$ is yet another harmonic function. The metric
(\ref{3chargemetric}) has
\be
\sqrt{2}\,\omega_0={Q\over 8}\,{\g_1^2-\g_2^2 + {4({\tilde r}_c-{\tilde
r})\over\eta}
-4({\tilde r}_c-{\tilde r})\Bigl(1+{\g_1\g_2(\g_1+\g_2)\over 2
(\g_1\,{\tilde r}+\g_2\,{\tilde r}_c)}\Bigr)
\over\g_1\,{\tilde r}+\g_2\,{\tilde r}_c}
\ee
and thus
\be
\sqrt{2}\,H_6={Q\over 8(\g_1+\g_2)}\,\Bigl({\g_1^2\over {\tilde
r}}-{\g_2^2\over{\tilde r}_c}\Bigr)
\ee

We thus explicitly identify the six harmonic functions $H_i$
($i=1,\ldots,6$) that describe  3-three charge solutions
(\ref{3chargemetric}). Any
choice of harmonic functions gives a solution to the equations of motion
(\ref{beta}-\ref{f}), but generic solutions constructed this
way will not be true bound states of the 3-charge system.
    In particular if we
superpose the harmonic functions $H_i$ corresponding to different
values of the
parameters $\g_1,\g_2$ to generate new geometries then we find
pathologies at the $f=0$ surface for instance, which we would not
expect for an actual 3-charge bound state.

\end{document}